\documentclass[prl,amsmath,amssymb,twocolumn,superscriptaddress]{revtex4-1}
\usepackage{amsmath,amssymb}
\usepackage[usenames]{color}
\usepackage{amssymb}
\usepackage{grffile}
\usepackage[pdftex]{graphicx}
\usepackage{amsmath, amstext, amssymb, amsfonts, amsxtra}
\usepackage{textcomp}
\usepackage{xspace}
\usepackage{bbm}
\usepackage{bm}
\newcommand{\be}{\begin{equation}}
\newcommand{\ee}{\end{equation}}

\newcommand{\xiamen}{Department of Physics and Key Laboratory of Low Dimensional
Condensed Matter Physics (Department of Education of Fujian Province), Xiamen
University, Xiamen 361005, Fujian, China}
\newcommand{\como}{Center for Nonlinear and Complex Systems, Dipartimento di
Scienza e Alta Tecnologia, Universit\`a degli Studi dell'Insubria,
via Valleggio 11, 22100 Como, Italy}
\newcommand{\infn}{Istituto Nazionale di Fisica Nucleare, Sezione di Milano,
via Celoria 16, 20133 Milano, Italy}
\newcommand{\brazil}{International Institute of Physics, Federal University
of Rio Grande do Norte, Campus Universit\'ario - Lagoa Nova, CP. 1613, Natal,
Rio Grande Do Norte 59078-970, Brazil}
\newcommand{\NEST}{NEST, Istituto Nanoscienze-CNR, I-56126 Pisa, Italy}

\begin{document}

\title{Inverse Currents in Hamiltonian Coupled Transport}

\author{Jiao Wang}
\affiliation{\xiamen}
\author{Giulio Casati}
\affiliation{\como}
\affiliation{\brazil}
\author{Giuliano Benenti}
\affiliation{\como}
\affiliation{\infn}
\affiliation{\NEST}

\begin{abstract}
The occurrence of an inverse current, where the sign of the
induced current is opposite to the applied force, is a highly counterintuitive
phenomenon. We show that inverse currents in coupled transport (ICC) of energy
and particle can occur in a one-dimensional interacting Hamiltonian system
when its equilibrium state is perturbed by coupled thermodynamic forces. This
seemingly paradoxical result is possible due to the self-organization occurring
in the system in response to the applied forces.
\end{abstract}

\maketitle

In the study of particle transport, inverse particle current, denoted as
absolute negative mobility (ANM), is arguably the most counterintuitive
transport phenomenon, in that a system responds to an applied static force by
generating a current against that force. It has been pointed out that ANM cannot
take place around a thermal equilibrium state~\cite{Vandenbroeck01, Hanggi02},
otherwise it could be exploited to construct a perpetuum mobile of the second
kind, with a single heat bath performing work. In nonequilibrium systems, however,
there is no fundamental law that forbids ANM, and indeed ANM has been investigated
in a variety of nonequilibrium setups, e.g., in relation to particle
separation~\cite{Hanggi12,Hanggi19}, self-propulsion~\cite{Hanggi14},
tracer dynamics in a steady laminar flow~\cite{Vulpiani16}, and also
experimentally in semiconductor superlattices~\cite{Keay95}, microfluidic
systems~\cite{Anselmetti05}, and Josephson junctions~\cite{Nagel08},
subject to ac electric fields.

The above argument of a perpetuum mobile only applies when there is a
single flow in response to a single driving force acting on the system.
It does not rule out the possibility of inverse currents
in coupled transport (ICC). Indeed, the entropy  production rate
$\dot{S}=J_1\mathcal{F}_1+J_2\mathcal{F}_2$, with $\mathcal{F}_i>0$ ($i=1,2$)
thermodynamic forces and $J_i$ conjugated currents, can be positive even though
one of the two induced currents has sign opposite to both forces (say, $J_1>0$
and $J_2<0$).
Nevertheless, the possibility that ICC can take place by perturbing an
equilibrium state by means of two forces, though not in contradiction
with thermodynamics, appears highly counterintuitive. Indeed, it would
imply that a system at equilibrium exposed to two thermodynamic forces,
could, under appropriate conditions, exhibit ICC \emph{against both forces}.
Actually, such a possibility has been recently shown to
occur in an abstract stochastic model~\cite{Mukamel}. More precisely,
it was obtained for a stochastic dynamics, with a tracer particle subject
to two driving forces and moving on a discrete ring populated by neutral
particles, which in turn obey a symmetric exclusion process. However,
such type of ICC has not been shown possible in any physical system.
In particular, in the model of hard Brownian disks in a narrow planar chain,
of which the above stochastic model serves as a toy model, ICC was not
found~\cite{Mukamel}. It raises the basic question if, besides
abstract stochastic models, the ICC phenomenon can take place in
a purely dynamical, Hamiltonian system when perturbing its equilibrium state.

In this Letter we give a positive answer to this question by considering
a one-dimensional (1D), two masses, interacting gas model. It is found that
by perturbing its equilibrium state with biases in temperature and chemical
potential, it is possible to have one flow (either particle or energy)
\emph{against both biases}. This seemingly
paradoxical result is possible due to a negative Onsager
cross coefficient for thermodiffusion, which in turn is rooted in the
surprising property of our model of adapting its structure in response
to external gradients, with separation of the two species of particles
of different masses~\cite{PRL2017}.

\begin{figure}[!]
\includegraphics[width=8.5cm]{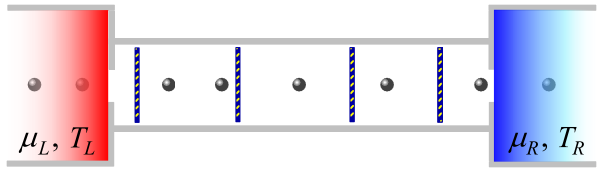}
\caption{Schematic drawing of the 1D, two mass, interacting gas model.
It consists of a diatomic gas of particles which, for visualization purposes,
we represent by bullets and rods, respectively.
The system is coupled to two reservoirs at its two ends, with which the rods
exchange energy while the bullets exchange both energy and particles.}
\label{fig1}
\end{figure}

Our model can be viewed as a classical version of the Lieb-Liniger
model~\cite{LL} for a diatomic gas. A schematic drawing is provided in
Fig.~\ref{fig1}, where the two species of particles of mass $\mathcal{M}_1$
and $\mathcal{M}_2$ are denoted as bullets and rods, respectively, for
visualization purposes. The masses are confined to move in a 1D box of
length $L$, with the Hamiltonian
\begin{equation}
H=\sum_{i} \frac{p_i^2}{2m_{i}}+\sum_{i<j}V(x_{i}-x_{j}).
\label{eq1}
\end{equation}
Here $m_i\in\{\mathcal{M}_1,\mathcal{M}_2\}$, $p_i$, and $x_i$ are,
respectively, the mass, momentum, and position of the $i$th particle,
and the potential $V(x)=h$ for $x \le |r|$ and $V(x)=0$ otherwise,
with $h\ge 0$ being the potential barrier.

%%%%%%%%%%%%%%%%%
%%%%%%%%%%%%%%%%%

In this work we consider the limit $r\to 0$ where all particles move freely
except when two particles meet. At such a moment, the two particles simply pass
through each other if their total energy in the frame of the center of mass
is larger than the potential barrier $h$; otherwise they collide elastically.
Note that for $h=0$, interactions vanish and the system is integrable.
For $h=\infty$, the system reduces to the nonintegrable, hard-core two-mass
gas~\cite{Giulio86}, an intensively studied paradigm for the 1D transport problem.
Interestingly, for a finite nonzero value of $h$, the system's transport behavior
is very rich~\cite{Kurchan, PRL2017}. In particular, as shown in~Ref.~\cite{PRL2017},
when the system is brought into contact with two heat baths at different temperatures,
it may undergo a nonequilibrium phase transition as the temperature difference
increases.

In the following we investigate a different setup where also the particle current
plays a role (see Fig.~\ref{fig1}): two reservoirs of one kind of particles,
say, the ``bullets'', are introduced to replace the heat baths. The reservoirs
are modeled as infinite 1D ideal gases~\cite{reservoir}, which are allowed to
exchange both energy and (bullet) particles with the system; i.e., when a bullet
particle hits a reservoir, it will be absorbed; meanwhile, the reservoirs
inject bullet particles into the system randomly, with rates $\gamma_\alpha$
($\alpha=L, R$) determined by their temperatures $T_\alpha$ and chemical
potentials $\mu_\alpha$ as~\cite{rate}

\begin{equation}
\gamma_{\alpha}=\frac{\rho_0}{\sqrt{2\pi \mathcal{M}_1 \beta_0}}
\frac{\beta_0}{\beta_\alpha}e^{\mu_\alpha \beta_\alpha-\mu_0 \beta_0}.
\label{eq3}
\end{equation}
Here, $\beta_0=1/(k_B T_0)$ with $k_B$ the Boltzmann constant, and $T_0$,
$\rho_0$, and $\mu_0$ are, respectively, the temperature, particle number
density, and chemical potential of a reference state. An injected particle
has a random velocity sampled from the distribution~\cite{bath}
\begin{equation}
P_{\alpha}(v,\mathcal{M}_1)={\mathcal{M}_1|v|\beta_\alpha}e^{-{\mathcal{M}_1
v^2\beta_\alpha}/2},
\label{eq2}
\end{equation}
and the time interval between two neighboring injections of a reservoir obeys
the Poisson distribution
$\tilde P_{\alpha} (t)=\gamma_{\alpha} e^{-\gamma_{\alpha} t}$.
Instead, the rod particles only exchange energy with the reservoirs; i.e., when
a rod particle hits a reservoir, it will be reflected back with a new velocity
randomly chosen from $P_{\alpha}(v,\mathcal{M}_2)$ ~\cite{note1}.
The total number of rod particles is thus conserved.

The system is subject to the thermodynamic forces
$\mathcal{F}_\rho=\mu_L\beta_L-\mu_R\beta_R$ and $\mathcal{F}_u=\beta_R-\beta_L$,
conjugated to the particle and energy currents, $J_\rho$ and $J_u$, respectively.
We set $\mathcal{F}_\rho>0$ and $\mathcal{F}_u>0$, so that a negative
current signals ICC. Note that the ICC phenomenon should not be confused with
thermodiffusion, where the two thermodynamic forces have opposite sign instead
and, for instance, the motion of particles against a chemical potential
difference is possible thanks to a temperature difference.

In our numerical simulations, we set
$T_L=T+\Delta T/2$, $\mu_L=\mu+\Delta \mu/2$, $T_R=T-\Delta T/2$, and
$\mu_R=\mu-\Delta \mu/2$. We focus on the illustrative example with $T=1$ and
$\mu=1.5$, but ICC has been verified to be independent of this particular choice.
Other parameter values are: $k_B=1$, $h=1$ throughout unless explicitly stated
otherwise, $\mathcal{M}_1=1$, and $\mathcal{M}_2=0.5$ (as explained below, to
have ICC it is crucial that $\mathcal{M}_1>\mathcal{M}_2$, that is, the particles
exchange with the reservoirs are the heavier ones). The number of rods is set to
be half of the expected particle number of a 1D ideal gas at the equilibrium
state  with given $T$ and $\mu$; i.e., $N_{\mathcal{M}_2} = \rho L/2$ with
$\rho=\rho_0(\sqrt{\beta_0/\beta})e^{\beta\mu-\beta_0\mu_0}$. (For
the reference state, $\rho_0=1$, $T_0=1$, and $\mu_0=0$.)
To evolve the system, an effective event-driven algorithm is
utilized~\cite{algorithm} to ensure the relative errors of all numerical
results are smaller than 0.5\%.

\begin{figure}[!t]
\includegraphics[width=8.0cm]{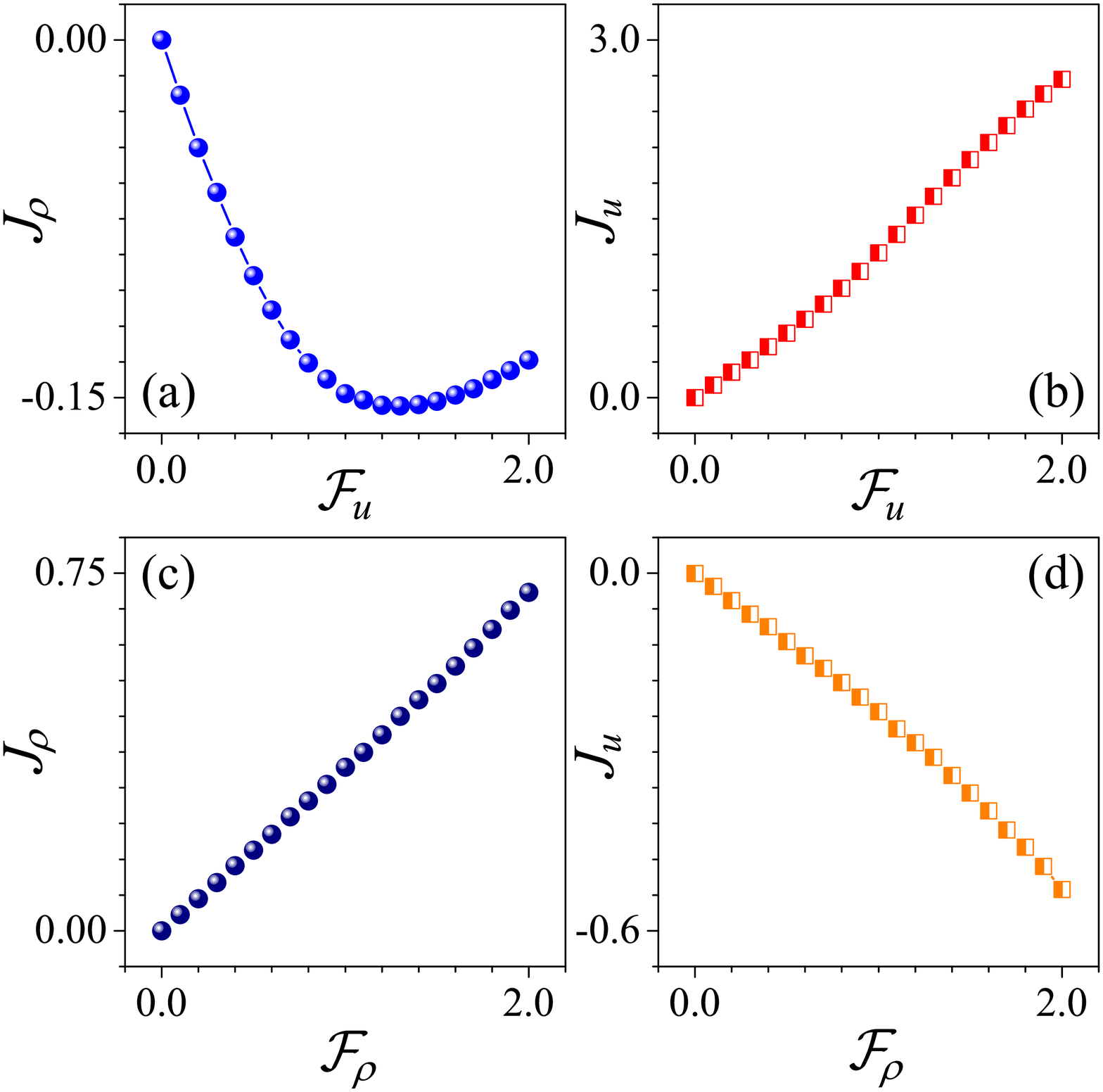}
\caption{The particle current $J_\rho$ (a) and the energy
current $J_u$ (b) as a function of the thermodynamic force $\mathcal{F}_u$
with $\mathcal{F}_\rho=0$. The negative particle current $J_\rho$ indicates ICC.
The system size is $L=20$. (c) and (d) are the same as (a) and (b), respectively,
but for the currents as a function of the thermodynamic force $\mathcal{F}_\rho$
(with $\mathcal{F}_u=0$). Here ICC is indicated by the negative energy current
$J_u$.}
\label{fig2}
\end{figure}

We start by seeking ICC in the simpler cases when only one thermodynamic
force acts. First, we set $\mathcal{F}_\rho=0$ to see how the currents depend
on the force $\mathcal{F}_u$. A typical result is shown in Figs.~\ref{fig2}(a)
and 2(b), where it is seen that while $J_u$ is positive and monotonically
increases with $\mathcal{F}_u$, $J_\rho$ is negative and decreases with
increasing $\mathcal{F}_u$, until it reaches a turning point. This is a clear
evidence of ICC, since over the whole range of $\mathcal{F}_u$ investigated
$J_\rho$ is negative, i.e., the particle current flows from the low to the
high temperature reservoir. Note that for small $\mathcal{F}_u$ ($<0.3$
in this case) both currents depend on $\mathcal{F}_u$ linearly, suggesting
that the system is in the linear response regime. Therefore, as indicated
by Fig.~\ref{fig2}(a), ICC can take place in both the linear response regime
and beyond.

%%%%%%%%%%%%%%%%%
%%%%%%%%%%%%%%%%%

Similarly, a parallel study reveals how the currents depend on
the thermodynamic force $\mathcal{F}_\rho$ with $\mathcal{F}_u=0$ [see
Figs.~\ref{fig2}(c) and 2(d)]. Again, ICC is observed, but in the energy current
$J_u$ instead, i.e., the energy flows from the reservoir of the lower chemical
potential to the opposite one (note that here the two reservoirs have the
same temperature since $\mathcal{F}_u=0$).

\begin{figure}[!t]
\vskip-0.3cm
\includegraphics[width=8.0cm]{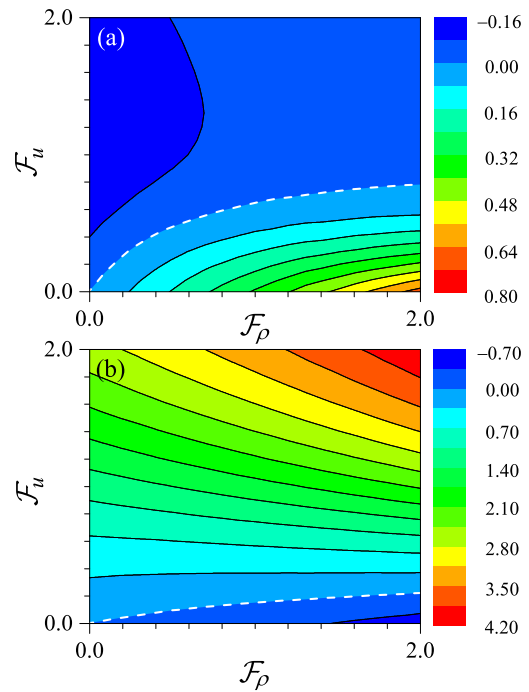}
\caption{The particle current $J_\rho$ (a) and the energy current $J_u$ (b) as
a function of the thermodynamic forces $\mathcal{F}_u$ and $\mathcal{F}_\rho$.
ICC occurs in $J_\rho$ when the thermodynamic forces fall in the area above
the white dashed curve in (a) while it occurs in $J_u$ when the thermodynamic
forces fall in the area below the white
dashed curve in (b). The system size is $L=20$.}
\label{fig3}
\end{figure}

In order to have an overall grasp of how the currents depend on the thermodynamic
forces, we thoroughly compute the currents for various $\mathcal{F}_\rho$ and
$\mathcal{F}_u$.  The results are summarized in Fig.~\ref{fig3}. An area of
($\mathcal{F}_\rho$, $\mathcal{F}_u$) for ICC in the particle current $J_\rho$
can be recognized in Fig.~\ref{fig3}(a) (above the white dashed line), while
the ICC in the energy current $J_u$ can be found in Fig.~\ref{fig3}(b) (below
the white dashed line). The two ICC areas do not overlap, as expected, since
simultaneous ICC in both currents would lead to a negative entropy production
rate, thus violating the second law of thermodynamics.

\begin{figure}[!t]
\includegraphics[width=8.0cm]{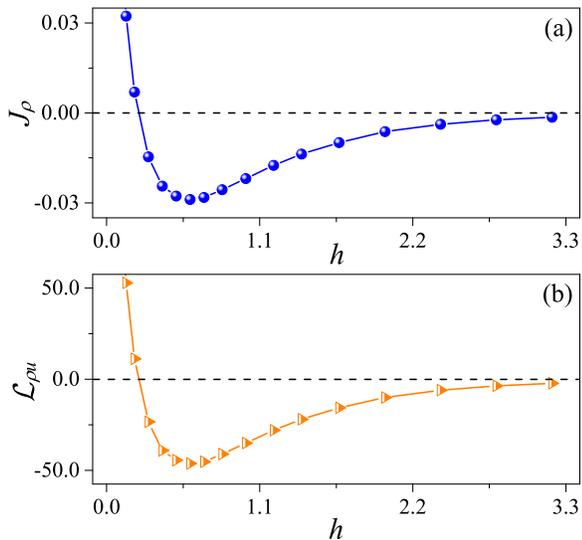}
\caption{The dependence on the interaction potential barrier $h$ of
(a) the particle current $J_\rho$ and (b) the Onsager cross coefficient
$\mathcal{L}_{\rho u}$. The thermodynamic forces are $\mathcal{F}_\rho=0$
and $\mathcal{F}_u=0.1$, respectively, and the system size is $L=160$. }
\label{fig4}
\vskip-0.15cm
\end{figure}

Next, we study the role the interactions play in generating ICC. To this end,
we investigate how currents depend on the potential barrier. For example,
in Fig.~\ref{fig4}(a), the results for $J_\rho$ with $\mathcal{F}_u>0$ and
$\mathcal{F}_\rho=0$ is given. It shows that as $h\to 0$, i.e., when the
interactions tend to vanish, $J_\rho$ increases and turns to be positive,
indicating that interactions are necessary to obtain ICC. On the other hand,
in the limit $h\to \infty$, when the particles tend to collide
elastically without passing through each other, the ICC current decays.
Therefore, allowing the particles passing through each other is a crucial
element of the interactions for inducing ICC in our system \cite{note2}.

%%%%%%%%%%%%%%%%%
%%%%%%%%%%%%%%%%%

In the linear response regime, the currents are related to the thermodynamic
forces as~\cite{Callen, Groot}
\begin{equation}
\left(
\begin{array}{c}
J_\rho\\
J_u
\end{array}
\right) = \left(
\begin{array}{cc}
\mathcal{L}_{\rho \rho} & \mathcal{L}_{\rho u} \\
\mathcal{L}_{u \rho} & \mathcal{L}_{u u}
\end{array}
\right) \left(
\begin{array}{c}
\mathcal{F}_\rho /L\\
\mathcal{F}_u /L
\end{array}
\right) ,
\label{eq:lresponse}
\end{equation}
where $\mathcal{L}_{ij}$ ($i,j=\rho, u$) are the Onsager kinetic coefficients.
For the general case where the currents are positive for positive forces,
$\mathcal{L}_{ij}>0$. However, thermodynamics does not forbid the
cross coefficients to be negative. As shown in Fig.~\ref{fig4}(b), our model
exhibits such an unusual feature, and indeed, it is when $\mathcal{L}_{\rho u}<0$
that ICC occurs in the linear response regime.

The fact that $\mathcal{L}_{\rho u}<0$ can be understood as follows.
If we set $\mathcal{F}_\rho=0$ and $\mathcal{F}_u>0$, the probability for
two particles to cross each other is higher when the light particle (a rod)
is closer to the hot end and the heavy particle (a bullet) is closer to the
cold end. In this case, the relative velocity of the two particles is on
average higher than in the opposite configuration, hence crossing is more
likely. This creates an unbalance in the particle density for the two species,
with the rods staying preferably closer to the cold side and the bullets close
to the hot side (such an unbalance develops to a phase separation in the
far-from-equilibrium regime). As only bullets exchange with the reservoir,
an average flow of bullets from the cold to the hot reservoir will form,
i.e., $J_\rho<0$. This in turn implies $\mathcal{L}_{\rho u}<0$
[see Eq.~(\ref{eq:lresponse})] in the linear response regime.

%%%%%%%%%%%%%%%%%
%%%%%%%%%%%%%%%%%

Now we state two properties of ICC in our model. First, numerical
data on the size dependence of ICC (see Sec.~I of the Supplemental
Material~\cite{SM}) show that, for fixed driving forces, ICC first becomes
stronger and then saturates with increasing system size, in both the linear
response regime and beyond. These data provide a clear indication for the
existence and relevance of ICC in our model in the limit of large system size.

\begin{figure}[!t]
\vskip-0.05cm
\includegraphics[width=8cm]{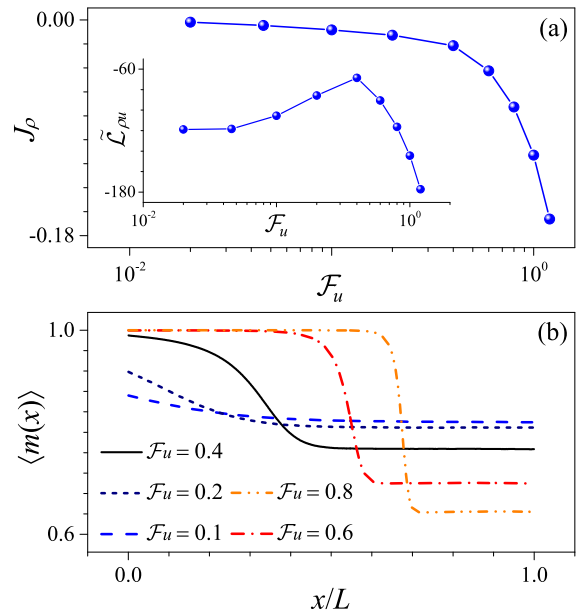}
\caption{(a) The particle current $J_\rho$ as a function of the thermodynamic
force $\mathcal{F}_u$ with $\mathcal{F}_\rho=0$. The system size is $L=1280$.
The inset shows $\tilde{\mathcal{L}}_{\rho u}\equiv J_\rho L / \mathcal{F}_u$.
(b) The average mass of all particles passing a given point $x$ along the
system. The five curves correspond to the middle five points shown in
panel (a).}
\label{fig5}
\vskip-0.15cm
\end{figure}

Second, for a large, fixed system size, a phase separation occurs when
driving forces exceed certain values, causing a dramatic enhancement
of the ICC effect. As an example, in Fig.~\ref{fig5}(a) is shown the particle
current $J_\rho$ driven by $\mathcal{F}_u$ with $\mathcal{F}_\rho=0$ for
a large system size $L=1280$. It can be seen that, below a turning point
$\mathcal{F}_u^{(c)}\approx 0.4$, $|J_{\rho}|$ increases with $\mathcal{F}_u$
slowly. Instead, for $\mathcal{F}_u > \mathcal{F}_u^{(c)}$, when the phase
separation takes place (see below), $|J_{\rho}|$ increases with $\mathcal{F}_u$
sharply. This abrupt change is even more evident in the rescaled particle
current $\tilde{\mathcal{L}}_{\rho u} \equiv J_\rho L/\mathcal{F}_u$ [the
inset of Fig.~\ref{fig5}(a)]. This quantity, which tends to
$\mathcal{L}_{\rho u}$ in the linear response regime $\mathcal{F}_u \to 0$,
first decays (in absolute value) for $\mathcal{F}_u < \mathcal{F}_u^{(c)}$
and then rapidly increases when $\mathcal{F}_u > \mathcal{F}_u^{(c)}$.

%%%%%%%%%%%%%%%%%
%%%%%%%%%%%%%%%%%

To illustrate the phase separation at $\mathcal{F}_u=\mathcal{F}_u^{(c)}
\approx 0.4$, we show in Fig.~\ref{fig5}(b) the averaged particle mass,
$\langle m(x) \rangle$, for all the particles that pass a certain position
$x$. We can see that, for small driving force, $\langle m(x) \rangle$ is
almost uniform except that at the left end it is a bit higher. Hence
the two types of particles are overall uniformly mixed. As the driving force
is strengthened, the left end of the curve lifts higher, suggesting that bullets
(of mass $\mathcal{M}_1=1$) tend to accumulate at the left end and rods (of mass
$\mathcal{M}_2=0.5$) are pushed to the right. When the driving force reaches
the critical value, the left end is exclusively occupied by the bullets. For
even stronger driving force, a domain of pure bullet particles emerges at the
left end, and the rods are brought out of contact with the left reservoir
completely, indicating that a critical transition in the system's structure has
occurred. The change in structure is a self-organization behavior to adaptively
respond to the external forces, which recalls that observed in the system's
heat conduction behavior~\cite{PRL2017}. We also note that ANM was observed
in Ref.~\cite{Vandenbroeck01} for Brownian particles, as a consequence of
particle-particle attractive interactions and of a kind of self-organization
in the system. While in Ref.~\cite{Vandenbroeck01} the dynamics is ruled by a
Markovian master equation and only the particle flow is considered; in our
case we consider coupled  flows for a Hamiltonian dynamics.

In summary, we have shown that ICC can take place in a 1D, interacting
Hamiltonian system, in either the particle or the energy current. The
effect is observed both within linear response and beyond and is rooted in
the self-organization in response to the applied forces, up to complete phase
separation in the far-from-equilibrium regime, where ICC is greatly enhanced.
Our results raise the question of what the general conditions for ICC are and
if other mechanisms different from self-organization exist. It might be also
interesting to explore possible new effects that negative Onsager
cross coefficients may induce~\cite{Lepri_note,Lepri2012}, of which those on
thermoelectricity~\cite{Benenti2017, Benenti2013, Benenti2014,Chen2015, Luo2018}
might be of particular interest~\cite{note3}. For instance, one could, in principle,
design a thermoelectric circuit~\cite{Benenti2017} with only one kind of electric
carrier rather than alternating $p$- and $n$-doped semiconductors as in a
thermocouple. The opposite response to a temperature difference could be obtained
by alternating a channel with standard response to applied thermodynamic forces
and a channel that exhibits ICC.

We acknowledge support by the NSFC (Grants No. 11535011 and No. 11335006) and
by the INFN through the project QUANTUM. The computational resources utilized
in this study were provided by the Shanghai Supercomputer Center.

\subsection{Supplemental Material: Inverse Currents in Hamiltonian Coupled Transport}

Here we provide more analysis of ICC in our model. For all the numerical
results presented, the parameters adopted in simulations are the same as
in the main text.

\subsection{I. Dependence on system size}

\begin{figure}[!b]
\vskip-0.3cm
\includegraphics[width=8.4cm]{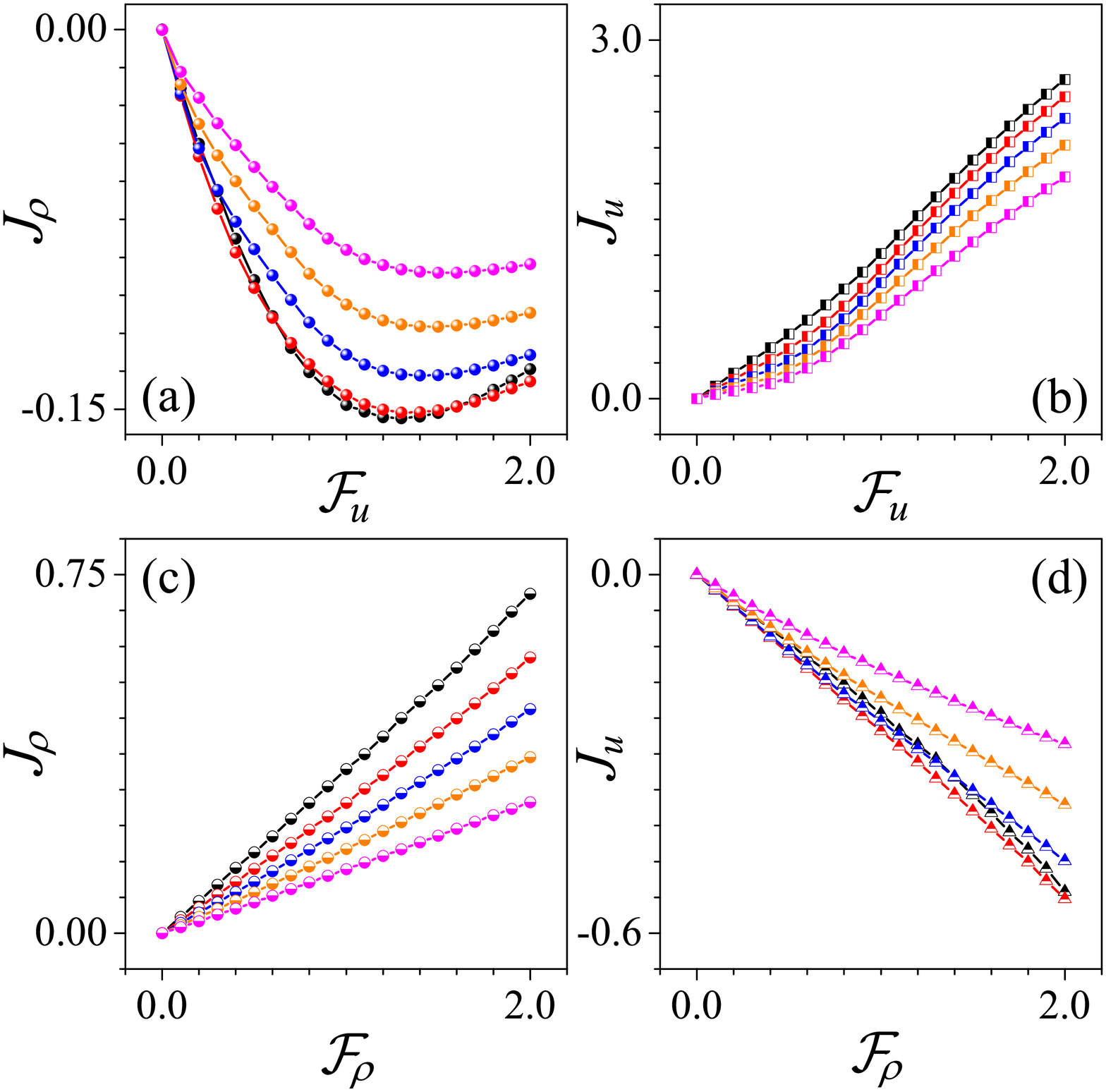}
\vskip-0.3cm
\caption{The same as Fig.~2 in the main text, but for five different system
sizes. In each panel, the black, red, blue, orange, and magenta curve are
for, respectively, the system size of $L=20$, 40, 80, 160, and 320.}
\label{figS1}
\end{figure}

\begin{figure}[!b]
\vskip-0.3cm
\includegraphics[width=8.4cm]{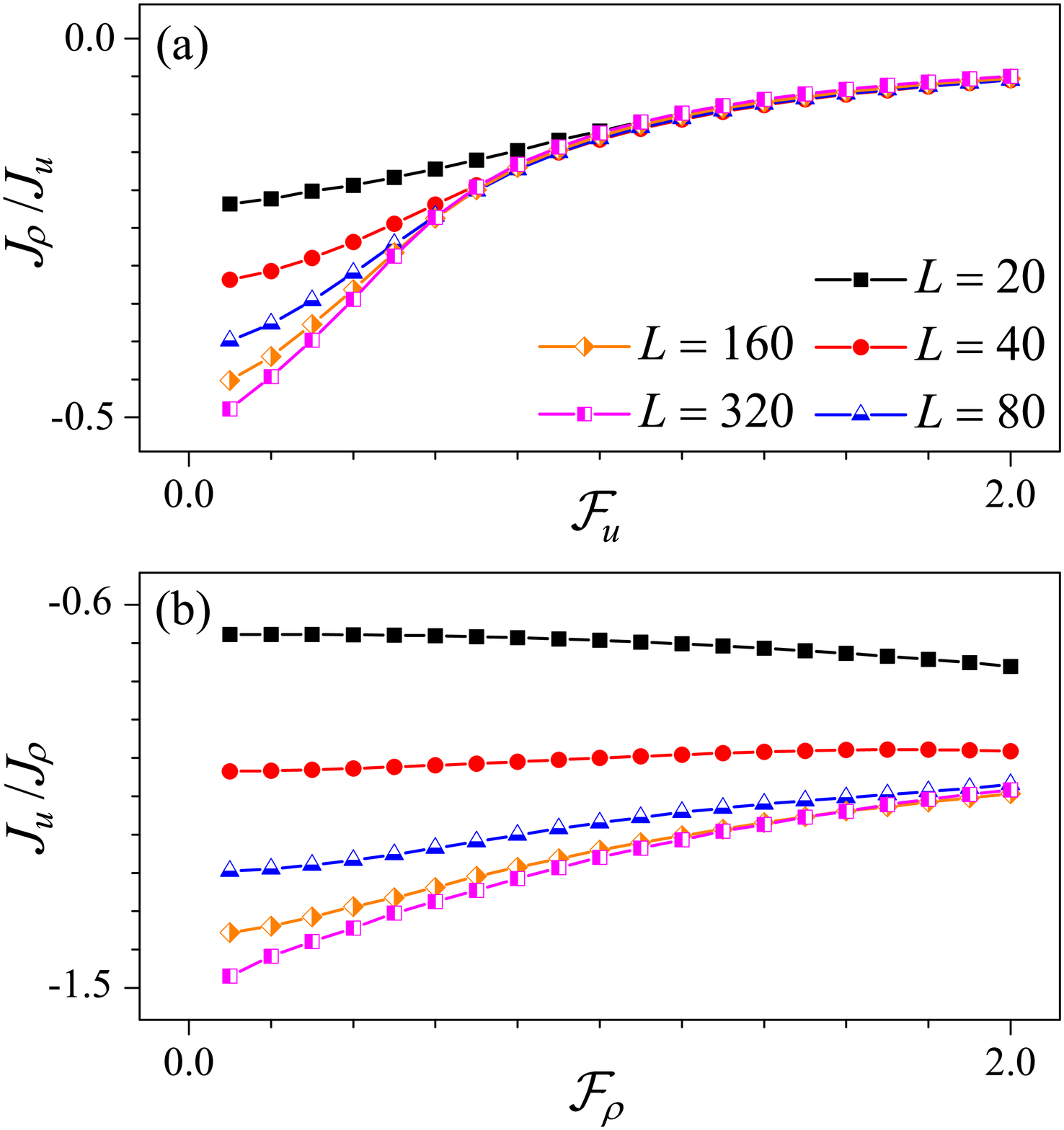}
\vskip-0.3cm
\caption{(a) Ratio $J_\rho/J_u$ versus the thermodynamic force
$\mathcal{F}_u$ (at $\mathcal{F}_\rho=0$) and (b) $J_u/J_\rho$ versus
$\mathcal{F}_\rho$ (at $\mathcal{F}_u=0$) for different system sizes.
Legends in panel (a) also apply to panel (b).}
\label{figS2}
\end{figure}

In Fig.~\ref{figS1}, the dependence of the currents $J_\rho$ and $J_u$ on
the thermodynamic forces $\mathcal{F}_\rho$ and $\mathcal{F}_u$ is shown
for various system sizes. It can be seen that, in general, the currents
drop as the system size increases. But, however, the relevance of ICC
is not decreasing. Indeed, as shown in Fig.~\ref{figS2}, at a given force
$\mathcal{F}_u$ or $\mathcal{F}_\rho$, as the system size increases, both
the negative ratio $J_\rho/J_u$  (at $\mathcal{F}_\rho=0$) and $J_u/J_\rho$
(at $\mathcal{F}_u=0$) would increase first (in absolute value) and then
tend to saturate.

%%%%%%%%%%%%%%%%%
%%%%%%%%%%%%%%%%%

%%%%%%%%%%%%%%%%%
%%%%%%%%%%%%%%%%%

\subsection{II. Thermoelectric implications}

In the linear response regime, the currents are related to the thermodynamic
forces as ~\cite{Callen,Groot}
\begin{equation}
\left(
\begin{array}{c}
J_\rho\\
J_u
\end{array}
\right) = \left(
\begin{array}{cc}
\mathcal{L}_{\rho \rho} & \mathcal{L}_{\rho u} \\
\mathcal{L}_{u \rho} & \mathcal{L}_{u u}
\end{array}
\right) \left(
\begin{array}{c}
\mathcal{F}_\rho /L\\
\mathcal{F}_u /L
\end{array}
\right) ,
\label{eq:lresponse}
\end{equation}
where $\mathcal{L}_{ij}$ ($i,j=\rho, u$) are the Onsager kinetic coefficients.

In our system, if the bullet particles are charge carriers, then the electric
conductivity $\sigma$, the thermal conductivity $\kappa$, and the Seebeck
coefficient $S$ of the system are connected with $\mathcal{L}_{ij}$ ($i,j=\rho, u$)
as
\begin{equation}
\sigma=\frac{e^2}{T}\,\mathcal{L}_{\rho\rho},
~\kappa=\frac{1}{T^2}\frac{\det\mathcal{L}}{\mathcal{L}_{\rho\rho}},
~S=\frac{1}{eT}\left(\frac{\mathcal{L}_{\rho u}}{\mathcal{L}_{\rho\rho}}-\mu\right),
\end{equation}
where $e$ is the charge of a bullet particle and ${\det\mathcal{L}}$ denotes
the determinant of the matrix of Onsager kinetic coefficients. The thermoelectric
figure of merit $ZT$ can be expressed in terms of these transport coefficients
as~\cite{Benenti2017}
\begin{equation}
ZT=\frac{\sigma S^2}{\kappa}\,T.
\end{equation}
Thermodynamics imposes $ZT\ge 0$, with the efficiency of heat to work conversion
$\eta=0$ when $ZT=0$ and $\eta\to\eta_C$ when $ZT\to\infty$, $\eta_C$ being the
Carnot efficiency.

%%%%%%%%%%%%%%%%%
\begin{figure}
\includegraphics[width=8.4cm]{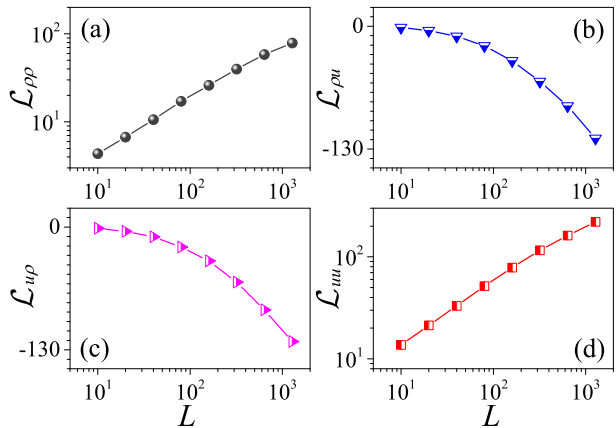}
\caption{The dependence of the Onsager kinetic coefficients on the system size.
The Onsager kinetic coefficients are evaluated based on Eq.~(\ref{eq:lresponse})
by computing the currents with ($\mathcal{F}_\rho$, $\mathcal{F}_u$) being
(0, 0.04) and (0.04, 0), respectively.}
\label{figS3}
\end{figure}
%%%%%%%%%%%%%%%%%

%%%%%%%%%%%%%%%%%
\begin{figure}
\includegraphics[width=8.4cm]{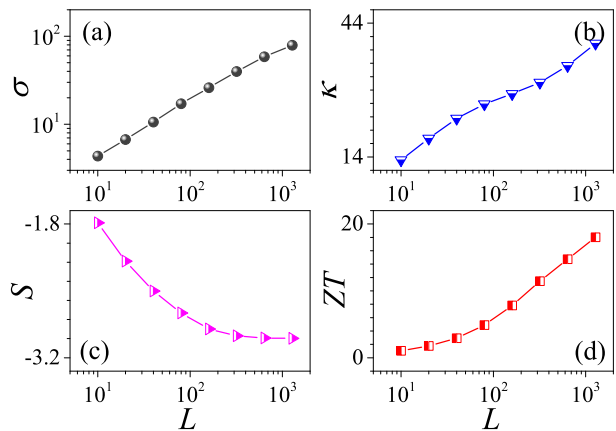}
\caption{The dependence of the transport coefficients and $ZT$ on the system
size in the linear response regime based on the computed Onsager kinetic
coefficients presented in Fig.~\ref{figS3}.}
\label{figS4}
\vskip-0.2cm
\end{figure}
%%%%%%%%%%%%%%%%%

In Fig.~\ref{figS3}, the system size dependence of the Onsager kinetic coefficients
is provided. It can be seen that the absolute value of all Onsager coefficients
increases with the system size. Moreover, the cross-coefficient $\mathcal{L}_{\rho u}$
is negative ($\mathcal{L}_{u \rho}=\mathcal{L}_{\rho u}$ due to the Onsager
reciprocal relations~\cite{Callen}). For simulations, the charge $e$ is set to be
unity throughout.

The size dependence of the transport coefficients and $ZT$ is shown in
Fig.~\ref{figS4}. It can be seen that while the Seebeck coefficient is negative
and saturates to a certain value, the electric conductivities $\sigma$ and
the heat conductivity $\kappa$ keep growing, but at different rates so that
$ZT$ increases monotonically. Extrapolation of these results to the thermodynamic
limit would imply that the Carnot efficiency would be achieved in that limit,
similar to other momentum conserving systems~\cite{Benenti2013, Benenti2014,
Chen2015, Luo2018}.

%%%%%%%%%%%%%%%%%%%%%%%%%%%%%%%%%%%%%%%%%%%%
%bibliography


\begin{thebibliography}{100}
\bibitem{note1}
The existence of ICC does not depend on the coupling between the rod particles
and the reservoirs. For example, ICC is verified to exist even if the rod particles
are insulated from the reserviors by setting their reflected velocity unchanged
but the direction.
\bibitem{note2}
This finding
is also supported by the results of ICC current $J_u$ when $\mathcal{F}_u=0$
but $\mathcal{F}_\rho>0$ and other cases where both thermodynamic forces are
nonzero.
\end{thebibliography}
\end{document}